\documentclass[showpacs,preprint,amsmath,PRB]{revtex4}
\usepackage{graphicx}% Include figure files
\usepackage{dcolumn}% Align table columns on decimal point
\usepackage{bm}% bold math

\begin{document}
%\twocolumn[\hsize\textwidth\linewidth\hsize\csname @twocolumnfalse\endcsname}

\title{On the transfer matrix method and WKB approximation for Schr\"{o}dinger equation with position-dependent effective mass}

\author{C F Huang$^{1,2}$, S D Chao$^{1,3,\ast}$, D R Hang$^{4,5,\ast}$, and Y C Lee$^{6}$ }

\affiliation{$^{1}$Department of Physics, National Taiwan University, Taipei, Taiwan, R. O. C.} 
\affiliation{$^{2}$National Measurement Laboratory, Center for Measurement Standards, Industrial Technology Research Institute, Hsinchu, Taiwan 300, R. O. C.}
\affiliation{$^{3}$ Institute of Applied Mechanics,  National Taiwan University, Taipei, Taiwan, R. O. C.}
\affiliation{$^{4}$ Department of Materials Science and Optoelectronic Engineering, National Sun Yat-sen University, Kaohsiung 804, Taiwan, R.O.C.}
\affiliation{$^{5}$ Center for Nanoscience and Nanotechnology, National Sun Yat-sen University, Kaohsiung 804, Taiwan, R.O.C.}
\affiliation{$^{6}$Institute of Materials Science and Engineering, National Sun Yat-sen University, Kaohsiung 804, Taiwan, R.O.C.}

\date{\today}% It is always \today, today,

\begin{abstract}
We have obtained a set of coupled differential equations from the continuous limit of the transfer matrix method. Decoupling such a set of equations yields an extension to the Wentzel-Kramers-Brillouin (WKB) approximation for the Schr\"{o}dinger equation with the position-dependent effective mass (PDEM). In the classically allowed region, the decoupling is to ignore the reflection resulting from the variations of both the potential and effective mass. By considering an infinite-well example with the PDEM, it is shown that the extended WKB approximation can provide not only an estimation to eigenenergies, but also an analytic form to approximate wavefunctions. \newline
* Corresponding authors: sdchao@spring.iam.ntu.edu.tw (S. D. Chao); drhang@mail.nsysu.edu.tw (D. R. Hang) 
\end{abstract}

\pacs{03.65.Ge}

\maketitle

\section{Introduction}

The one-dimensional Schr\"{o}dinger equation with the position-dependent effective mass (PDEM) has been introduced for many microstructures of current interest. \cite{Ou,Serea,Einevoll,Miranda} To understand such an equation analytically, exactly solvable examples have been discussed in the literature. \cite{Jiang,Alhaidari,Gonul} On the other hand, the transfer matrix method \cite{Forrest,Ozaydin,Hutchings,Yang1,Yang2} is a powerful numerical way to calculate the eigenvalues and eigenfunctions of the Schr\"{o}dinger equation with the PDEM. To perform this numerical method, the space is divided into a series of slabs so that the effective mass and potential can be approximated as constants in a slab. By considering the conservation of the flux, it has been suggested that we shall use the BenDaniel condition to match an eigenfunction at each boundary between two adjacent slabs. \cite{Forrest,Ozaydin,Hutchings,BenDaniel} In Ref. 1, Ou, Cao, and Shen derived a dispersion relation for the one-dimensional Schr\"{o}dinger equation with the PDEM by considering the analytic transfer matrix method. Such a dispersion relation contains an integral term of Wentzel-Kramers-Brillouin (WKB) type. It is well-known that WKB approximation \cite{Ou,Friedrich,Bjorken,Landau,Sakurai} provides important classical interpretations in the developement of quantum mechanics. Hence it is interesting to probe the validity of WKB approxmation after incorporating the variation of the effective mass.  

In this paper, a set of coupled differential equation is obtained by considering the continuous limit of the transfer matrix method. Decoupling such a set of equations, the approximate wavefunction of the analytic form 
\begin{eqnarray}
\psi (x) = \sqrt{ \frac{ m ^{\ast} (x) }{ k(x) } } exp [ \pm i S(x) / \hbar]
\end{eqnarray}
with  
\begin{eqnarray}
S(x) \equiv \int ^{x} k (x ^{\prime}) d x ^{\prime}  
\end{eqnarray}
is derived in section II when the effective mass $m ^{\ast} (x)$ depends on the position $x$. Here $V(x)$ is the potential, $\hbar$ equals Planck constant divided by $2 \pi$, $k(x) \equiv \sqrt{ 2 m ^{\ast} (x ) (E-V(x)) } / \hbar $, and $E$ is the eigenenergy. The phase $S(x)$ is of the same form as that in the well-known WKB approximation except that $m ^{\ast} (x)$ is position-dependent. Therefore, we can extend such an approximation for the one-dimensional Schr\"{o}dinger equation with the PDEM from the transfer matrix method. When the effective mass becomes constant in space, the approximate solution given by Eq. (1) is reduced to the conventional WKB-type function. As discussed in section III, the decoupling in the classically allowed region is to ignore the reflection due to the variations on the effective mass and/or potential. From our study, it becomes clear how to relate the numerical solution obtained from the transfer matrix method to the analytic WKB approximation. By considering an infinite-well example, it is shown that the extended WKB approximation can provide not only an estimation to eigenenergies, but also an analytic way to understand the eigenfunctions when the effective mass is position-dependent. The conclusion is made in section IV. 

\section{Transfer matrix method and approximate wavefunction of WKB-type}
When the effective mass $m^{\ast}$ depends on the position $x$, it has been shown that the one-dimensional Schr\"{o}dinger equation should be modified as \cite{Ou,Serea,Einevoll,Miranda}
\begin{eqnarray}
-\frac{d}{dx} \frac{\hbar ^{2}}{2 m ^{\ast} (x)} \frac{d}{dx} \psi (x) + V(x) \psi (x) = E \psi (x). 
\end{eqnarray}
In this paper, we assume that $m ^{\ast} (x)>0$. In addition, let $m ^{\ast} (x)$ and $V(x)$ be analytic functions. For convenience, in this section we only consider the classically allowed region, where
\begin{eqnarray}
E-V(x)>0
\end{eqnarray}
at any point $x$. To extend WKB approximation, the equation  
\begin{eqnarray}
\frac{d}{dx} \left[ \begin{array}{c} t(x) \\ r(x) \end{array} \right] = {\bf \Gamma} (x) \left[ \begin{array}{c} t(x) \\ r(x) \end{array} \right]
\end{eqnarray}
will be derived in this section by considering the continuous limit of the transfer matrix method. Here the position-dependent matrix 
\begin{equation}
{\bf \Gamma} (x) \equiv \left[ 
\begin{array}{c}
-ixk^{\prime }(x)-\frac{m^{\ast }(x)}{2k(x)}\left( \frac{k(x)}{m^{\ast }(x)}%
\right) ^{\prime },\text{ \ \ \ }\frac{m^{\ast }(x)}{2k(x)}\left( \frac{k(x)%
}{m^{\ast }(x)}\right) ^{\prime }e^{-2ixk(x)} \\ 
\frac{m^{\ast }(x)}{2k(x)}\left( \frac{k(x)}{m^{\ast }(x)}\right) ^{\prime
}e^{2ixk(x)},\text{ \ \ \ \ } ixk^{\prime }(x)-\frac{m^{\ast }(x)}{2k(x)}%
\left( \frac{k(x)}{m^{\ast }(x)}\right) ^{\prime }
\end{array}
\right], 
\end{equation}
$t(x)$ and $r(x)$ are two complex functions, and in this paper the notation $f ^{\prime} (x)$ presents the derivative of any function $f (x)$. With some calculations, we can see that Eq. (5) is equivalent to Eq. (3) if we set
\begin{eqnarray}
\psi (x) = t(x) e ^{ik(x)x} +r (x) e ^{-ik(x)x}.
\end{eqnarray} 
Then approximate solutions of WKB type are obtained by decoupling Eq. (5). Therefore, we can relate WKB-type solution to the transfer matrix method. 
     
The transfer matrix method is a powerful numerical approach to solve Eq. (3). \cite{Forrest,Ozaydin,Hutchings} To perform such a method, as shown in Fig. 1, we can divide the space into a series of thin slabs so that both the effective mass and potenital can be approximated as constants in each slab. For convenience, each slab is labelled by an integer $j$ sequently from the left to right. In the classical allowed region, the wavefunction $\psi (x)$ in a slab can be approximated as the linear superposition of two plane waves. Let $x _{j}$ be the center point of j-th  slab, and approximate the effective mass,  potential, and wave number as $m^{\ast} _{j}=m^{\ast}( x _{j} )$, $V _{j}=V(x _{j})$, and $k _{j}=k(x _{j})$ in such a lab, respectively. Then the wave function is taken as 
\begin{eqnarray}
\psi (x) = t _{j} e ^{ik _{j} x} +r _{j} e ^{-ik _{j} x} 
\end{eqnarray}
in the j-th slab. The coefficients $t _{j}$ and $r _{j}$ are the traveling components for two different directions. We shall consider the BenDaniel condition \cite{BenDaniel} to relate the coefficients of adjacent slabs. \cite{Lee,other} Based on such a condition, as shown in Appendix, the coefficients $t _{j+1}$ and $r _{j+1}$ are related to $t _{j}$ and $r _{j}$ by 
\begin{eqnarray}
\left[ \begin{array}{c} t_{j+1} \\ r_{j+1} \end{array} \right] = {\cal T}_{j} \left[ \begin{array}{c}
t_{j} \\ r_{j} \end{array}
\right] 
\end{eqnarray}
with the transfer matrix 
\begin{eqnarray}
{\cal T} _{j}=\left[ 
\begin{array}{c}
\frac{1}{2}(1+\frac{k_{j}}{m^{\ast }_{j}}\frac{ m^{\ast}_{j+1} }{ k_{j+1} }) e^{i(k_{j}-k_{j+1})y_{j}},\text{ \ \ \ }\frac{1}{2}(1-\frac{ k_{j} }{ m^{\ast}_{j}}\frac{m^{\ast }_{j+1}}{k_{j+1}})e^{-i(k_{j}+k_{j+1})y_{j}} \\ 
\frac{1}{2}(1-\frac{k_{j}}{m^{\ast }_{j}}\frac{m^{\ast }_{j+1}}{k_{j+1}})e^{i(k_{j}+k_{j+1})y_{j}},\text{\ \ \ \ \ }\frac{1}{2}(1+\frac{k_{j}}{m^{\ast}_{j}}\frac{m^{\ast }_{j+1}}{k_{j+1}})e^{-i(k_{j}-k_{j+1})y_{j}}
\end{array}
\right]
\end{eqnarray}
after some calculations. Here $y_{j}$ is the point separating the j-th and (j+1)-th slabs, so $x _{j+1} = (y _{j}+y _{j+1} )/2$. It is shown in Appendix that we can relate the transfer matrix ${\cal T}_{j}$ to ${\bf \Gamma (x)}$ by 
\begin{eqnarray}
{\cal T}_{j} = {\bf I} + {\bf \Gamma} (y_{j}) \Delta x_{j} + o(\Delta x_{j} ^{2}) 
\end{eqnarray}
with $\Delta x _{j} \equiv x _{j+1} - x _{j}$ and 
\begin{eqnarray}
{\bf I}\equiv \left[ \begin{array}{c} 1,\text{ \ \ \ \ }0 \\ 0,\text{ \ \ \ }\ 1 \end{array} \right].
\end{eqnarray}
Let $\Delta t _{j} \equiv t _{j+1} - t _{j}$ and $\Delta r _{j} \equiv r _{j+1} - r _{j}$. From Eqs. (9) and (11), we have the following set of equation, 
\begin{eqnarray}
\frac{1}{\Delta x _{j}}\left[ \begin{array}{c} \Delta t _{j} \\ \Delta r _{j} \end{array}
\right] ={\bf \Gamma}(y _{j})\left[ \begin{array}{c} t_{ x_{j} } \\ r_{ x_{j} } \end{array} \right] +o(\Delta x _{j}).
\end{eqnarray}
By shriking the widths of slabs so that $\Delta x _{j}$ approaches zero, the above equation can be reduced to Eq. (5) with the following correspondence
\begin{eqnarray}
\left[ \begin{array}{c} t _{j} \\ r _{j} \end{array} \right] 
\leftrightarrow \left[ \begin{array}{c} t (x_{j}) \\ r (x_{j}) \end{array} \right].
\end{eqnarray}
Thus we can obtain Eq. (5) by considering the continuous limit of transfer matrix method. From Eqs. (8) and (14), we can see why the wave function is determined by Eq. (7) after obtaining $r(x)$ and $t(x)$. 

In Eq. (5), the coefficients $t(x)$ and $r(x)$ are coupled to each other by the off-diagonal terms of $\Gamma (x)$. With some calculations, we can see from Eq. (11) that these terms comes from the off-diagoal terms of ${\cal T} _{j}$, which corresponds to the reflection due to the variations on potential and/or effective mass \cite{Forrest,Ozaydin,Hutchings}. If we ignore the reflection by neglecting the coupling bwtween $t(x)$ and $r(x)$, we have the following decoupled equations 
\begin{eqnarray}
\frac{d}{dx} t(x) \sim [ -i x k ^{\prime} (x) - \frac{m ^{\ast} (x)}{2k(x)} \left( \frac{k(x)}{ m ^{\ast} (x) }\right) ^{\prime} ] t(x)
\end{eqnarray}
\begin{eqnarray}
\frac{d}{dx} r(x) \sim [ i x k ^{\prime} (x) - \frac{m ^{\ast} (x)}{2k(x)} \left( \frac{k(x)}{ m ^{\ast} (x) }\right) ^{\prime} ] r(x).
\end{eqnarray}
 Solving the above two first-order differential equations, we have 
\begin{eqnarray}
t(x)e ^{ik(x)x} \sim c_{1} \sqrt{ \frac{m ^{\ast}(x)}{k(x)} } exp (iS(x)/ \hbar )
\end{eqnarray}
\begin{eqnarray}
r(x)e ^{-ik(x)x} \sim c_{2} \sqrt{ \frac{m ^{\ast}(x)}{k(x)} } exp (-iS(x) / \hbar).
\end{eqnarray}
Here $c _{1}$ and $c _{2}$ are two constants. From Eq. (7), the approximate solution is of the form 
\begin{eqnarray}
\psi (x) \sim c_{1} \sqrt{ \frac{m ^{\ast}(x)}{k(x)} } exp (iS(x)/ \hbar)+c_{2} \sqrt{ \frac{m ^{\ast}(x)}{k(x)} } exp (-iS(x) / \hbar).
\end{eqnarray}
We shall set $c _{1}$=0 ($c _{2}$=0) such that $t(x)$=0 ($r(x)$=0) for the traveling wave moving to right (left), and obtain the WKB-type function given by Eq. (1). Therefore, the analytic functions of WKB-type can be related to the numerical transfer matrix method by ignoring the reflection in the continuous limit. 
   
\section{Discussion}
In the last section, the approximate solution of WKB-type is derived for the classically allowed region by ignoring the off-diagonal terms of ${\bf \Gamma} (x)$ to decouple Eq. (5). With increasing eigenenergy $E$, the ignorance is reasonable as $k(x)>>1$ because the off-diagonal terms contain the factor $e ^{\pm 2 i k(x)x}$, which oscillates quickly under large $k(x)$. On the other hand, the diagonal terms of ${\bf \Gamma}(x)$ do not contain such an oscillating factor. Hence Eq. (1) provides a good approximation for the states with high energies as in the conventional WKB approximation. Because the transfer matrix ${\cal T}_{j}$ can be related to ${\bf \Gamma}(x)$ by Eq. (11), we can expect that the numerical solution obtained by the transfer matrix method can be reduced to the WKB-type function as $k(x)>>1$.

It has been shown that the dispersion relation of WKB-type can be used to estimate the eigenenergies of Schr\"{o}dinger equation with the PDEM. \cite{Ou} To further probe the extension of the WKB approximation to systems with PDEM, we consider the infinite quantum well where   
\begin{eqnarray}
\ m^{*}(x)= \frac{ m_{1} - m_{2} }{2a} x + \frac{ m_{1} + m_{2} }{2}
\end{eqnarray}
and $V(x)=0$ if $|x|<a$ while $V(x)=\infty$ if $|x| \geq a$. We do not need to consider how the phase changes at the turning points in this case, \cite{Ou,Friedrich} and just need to set eigenfunctions as zeros at $x=\pm a$. Hence we shall take $c_{1} = - c _{2} \equiv c$ in Eq. (19) to obtain the approximate eigenfunctions of WKB type
\begin{eqnarray}
\psi (x) \sim c \sqrt{ \frac{m ^{\ast}(x)}{k(x)} } sin (iS(x)/ \hbar) 
\end{eqnarray}
if we set $S(x) = \int _{-a} ^{x} \sqrt{ 2 m ^{*} (x ^{\prime}) (E- V(x ^{\prime})) } d x ^{\prime} $ and require $S(a)=n \pi$. Here $n$ is a nonnegative integer. Then we can obtain the WKB dispersion relation 
\begin{eqnarray}
E_{n}= \frac{9  n^{2} (m _{1} - m _{2}) ^{2} h^{2} }{128 (m _{1} ^{3/2} - m _{2} ^{3/2} ) ^{2} a ^{2} }  \ \text{with the integer} \ n = 1, 2, 3,...
\end{eqnarray}
On the other hand, we can reduce Eq. (3) as 
\begin{eqnarray}
\frac{d ^{2}}{dy ^{2}} u(y)= y u(y)
\end{eqnarray}
in such an example if we set 
\[
y=-(\frac{ (m _{1} - m _{2})E}{\hbar ^{2} a}) ^{1/3} (x + \frac{ m _{1} + m _{2} }{ m _{1} - m _{2}} a )
\]
and $u(y) =  \psi ^{\prime} (x) / m^{*} (x)$. The solution of the above equation is the linear combination of Airy functions, \cite{Sakurai,Airy} and thus it is easy to obtain the exact eigenvalues and wavefunctions. In Table 1, we compare the exact and approximate eigenenergies when $m_1 =0.1$ $m_0$, $m _{2}=0.2$ $m_0$, and $a=5$ $nm$. Here $m_{0}$ is the rest mass of electrons. We can see from table 1 that the errors of WKB approximation are reduced with increasing $n$. The error for the $n$=4 excited eigenlevel is only about $0.14 \%$, and we can see from Fig. 2 (a) that the corresponding WKB wavefunction is very close to the exact one. (In Fig. 2, both the approximate and exact wavefunctions are normalized such that $\int _{-a}  ^{a} |\psi (x)| ^2 dx =1$.) Therefore, the extended WKB approximation provides not only a good estimation to the eigenenergies, but also a good way to understand the wavefunctions. Figure 2 (b) shows the square of $n=4$ wavefunction. The oscillating amplitudes, in fact, are proportional to the factor $\sqrt{m^{*}(x)/k(x)}$ as expected from Eq. (21).                           

For the tunneling through a classically forbidden region where $E-V(x)<0$, it is known that we can obtain an estimation for the tunneling probability from WKB approximation if the (effective) mass is constant. With some calculations, we can obtain Eq. (5) with the same ${\bf \Gamma} (x)$ for the classically forbidden region from the continuous limit of the transfer matrix method. (It should be noted that $k(x)$ becomes imaginary.) In addition, the WKB-type approximate solution given by Eq. (1) can still be obtained by decoupling Eq. (5). In fact, it is shown in Ref. \cite{Alhaidari} that Eq. (3) can be transformed to the conventional Schr\"{o}dinger equation
\begin{eqnarray}
[\frac{d ^{2} }{ dy ^{2} } - \frac{2}{ \hbar ^{2} } ( {\cal V} (y) -E ) ] \phi (y)=0
\end{eqnarray}
if we set $y = \int \sqrt{ m ^{*} (x)} dx$, $\phi (y) = \psi (x) /  m ^{*1/4} (x) $, and ${\cal V}(y)= V(x) + F(x)$. Here 
\[
F(x)=-\frac{ \hbar ^{2} }{8 m ^{*}(x)}[\frac{\frac{d ^{2}}{dx ^{2}} m^{*}(x)}{m ^{*}(x)}-\frac{7}{4}(\frac{\frac{d}{dx} m^{*}(x)}{m^{*}(x)})^{2}]. 
\]
With some calculations, in fact, we can see that Eq. (1) can also be obtained from the conventional WKB-type function by considering such a transform after ignoring $F(x)$. The ignorance is reasonable when $m ^{*} (x)$ varies so slowly that its derivatives are small. Because WKB approximation may provide a suitable approximation to the above equation in the classically forbidden region, Eq. (1) also provides an extension to the WKB approximation when $E -V(x)<0$. 

Just as in the conventional WKB approximation, the reflection is neglected if we use Eq. (1) to approximate the wavefunction. If the effective mass or potential has sharp jumps at some interfaces, we cannot ignore the reflection and Eq. (1) does not give us a good approximate solution in the whole space. But we may apply Eq. (19) piecewisely to approximate the wavefunction. For example, consider a finite-well problem where the potential and effective mass have sharp jumps at two interfaces $A _{1}$ and $A _{2}$, as shown in Fig. 3. We can divide the space into three regions denoted by (I), (II), and (III), respectively, and approximate the wavefunction by Eq. (19) in each region. Then we just need to consider the BenDaniel condition at $A _{1}$ and $A _{2}$ to obtain the approximate solution in the whole space.

\section{Conclusion}
In this paper, we derive a set of first-order differential equations corresponding to the continuous limit of the transfer matrix method. Decoupling such a set of equations, an approximate wavefunction of WKB-type is obtained for the one-dimensional Schr\"{o}dinger equation with the position-dependent effective mass. In a classically allowed region, the decoupling is to ignore the reflection induced by the variations of the effective mass and potential. Such ignorance is reasonable in the high-energy limit. From our derivation, it is clear how to relate WKB approximation to the numerical solution obtained by the transfer matrix method when the effective mass is position-dependent. By considering an infinite-well example, it is shown that such an approximation can provide not only a simple estimation to eigenenergies, but also an analytic form to approximate the eigenfunctions.

\section*{Acknowledgements}
This work is partly supported by the National Science Council of the Republic of China under grant no: NSC 94-2112-M-110-009. D. R. Hang acknowledges financial supports from Aim for the Top University Plan and National Sun Yat-sen University, Kaohsiung, Taiwan. The work of S. D. Chao is supported by the National Science Council of the Republic of China.   

\section*{Appendix}
In the transfer matrix method, the BenDaniel condition is taken into account to match $\psi$ and $\psi ^{\prime} / m ^{\ast} $, the wavefunction and its derivative divided by the effective mass, at the boundary of each slab. At the boundary $y_{j}$ separating the j-th and (j+1)-th slabs, we have
\begin{eqnarray}
t_{j+1} e ^{i k_{j+1} y _{j}} + r_{j+1} e ^{ - i k_{j+1} y _{j}} = t_{j} e ^{i k_{j} y _{j}} + r_{j} e ^{i k_{j} y _{j}}
\end{eqnarray}
by matching $\psi$ and
\begin{eqnarray}
i \frac{ k _{j+1} }{ m ^{\ast} _{j+1} } t_{j+1} e ^{i k_{j+1} y _{j}} - i \frac{ k _{j+1} }{ m ^{\ast} _{j+1} } r_{j+1} e ^{ - i k_{j+1} y _{j}} = i \frac{ k _{j} }{ m ^{\ast} _{j} } t_{j} e ^{i k_{j} y _{j}} - i \frac{ k _{j} }{ m ^{\ast} _{j} } r_{j} e ^{i k_{j} y _{j}}
\end{eqnarray}
by matching $\psi ^{\prime} / m ^{ \ast }$ if $\psi$ is approximated by Eq. (8). We can obtain Eqs. (9) and (10) by relating $t _{j+1}$ and $r _{j+1}$ to $t_{j}$ and $r _{j}$ from the above two equations. 

By introducing the function $h(x) \equiv k(x) / m ^{\ast} (x)$, the factor 
\begin{eqnarray}
\frac{ k _{j} }{ m ^{\ast} _{j} }\frac{ m ^{\ast} _{j+1} }{ k _{j+1} } = \frac{ h( x _{j} ) }{ h (x _{j+1} ) } =  1 - \frac{ h ^{\prime} ( y _{j}) }{ h ( y _{j} ) } \Delta x + o ( \Delta x ^{2}).    
\end{eqnarray}             
Based on the above equation, we can obtain Eq. (11) as the first-order approximation with respect to $\Delta x$ since $e ^{ \pm i ( k _{j+1} -k _{j} ) y _{j}} \sim 1 \pm i y_{j} k ^{\prime} ( y _{j} ) \Delta x $ and $e ^{ \pm i ( k _{j+1} + k _{j} ) y _{j}} \sim e ^{\pm 2 i k (y _{j}) y_{j} }$.

Figure caption \newline
Fig.1 To perform the transfer matrix method, the space is divided into a series of slabs, in each of which both the potential and effective mass are approximated as constants. The points $x_{j}$ and $x_{j+1}$ present the centers of the j-th and (j+1)-th slabs, and $y _{j}$ is the point to separate these two slabs. \newline
Fig.2 (a) The red dash line and black dot line corresponds to the exact wavefunction and WKB approximate one. (b) The solid line is the square of the wavefunction and the dot line is the envelope function proportional to $m^{\ast}(x) ^{1/2}$. \newline
Fig.3 A finite-well problem where both the effective mass and potential have sharp jumps at the interfaces $A _{1}$ and $A _{2}$. The regions denoted by (I), (II) and (III) are divided by these two interfaces. 
\newline \newline
Table caption \newline
Table 1 The comparison of WKB and exact eigenvalues for the solvable example discussed in the text.    

\end{document}